\documentstyle[12pt]{article}
\begin{document}

\begin{center}
{\bf Particle production by the expanding thin-walled bubble}

\vspace{0.8cm} { Michael Maziashvili }

\vspace{0.5cm}\baselineskip=14pt

\vspace{0.5cm} \baselineskip=14pt {\it Department of Theoretical
Physics, Tbilisi State University, 0128 Tbilisi, Georgia}

\vspace{0.5cm}
\begin{abstract} Neglecting the effect of particle production at the
moment of bubble nucleation, the spectrum of created particles
during the bubble expansion is evaluated in the thin-wall
approximation. It is shown that the expanding thin-walled bubble
makes the dominant contribution to the particle production.
\end{abstract}

\end{center}

In this paper, as a step toward full understanding of the matter,
we extend our analysis \cite{Ma1} and investigate the particle
creation during the process of bubble growth, i.e. after the
bubble nucleation, in the thin wall approximation. It has been
shown that in the thin-wall approximation the particle production
at the moment of bubble nucleation is strongly suppressed
\cite{Ma1}. On the other hand, numerical calculations show that
the particle production by the thick-walled bubble is very
intensive \cite{Ma2}. The latter observation has led us to
consider the particle creation which occurs during the process of
bubble expansion in the thin-wall approximation. The question we
ask is the following: can one neglect in the thin wall case the
particle production after the bubble nucleation in comparison with
the number of particles produced during the nucleation process?
The expanding bubble is described by the $O(3,1)$ invariant
function which is an analytic continuation of the bounce solution
\cite{Co}. In the thin wall case, the bounce looks like a large
four-dimensional spherical bubble with a thin wall separating the
false vacuum without from the true vacuum within. As the bubble
expands, the wall traces out the hyperboloid. The bounce solution
provides the classical background with respect to which the
quantum fluctuations are defined. The second quantization picture
for the fluctuation field in process of false vacuum decay through
the barrier penetration is thoroughly examined in \cite{YTS,HSTY}.
Throughout this paper the signature for the Minkowskian space-time
metric is $(-,+,+,+)$ and $\hbar=c=1$ is assumed. For the
calculation of particle spectrum we use the standard formalism
developed in \cite{YTS,HSTY}.

The lagrangian density is
\begin{equation}{\cal L}=-\frac{1}{2}(\partial_{\mu}\phi)^2-U(\phi),\end{equation}
where
\begin{equation}\label{po}U(\phi)=\frac{\lambda}{2}\phi^2(\phi-2a)^2-\epsilon(\phi^3/2a^3-3\phi^4/16a^4),\end{equation}
and $a,~\lambda,~\epsilon$ are positive parameters. The potential
(\ref{po}) has a local minimum at $\phi_f=0,~U(0)=0,$ a global
minimum at $\phi_t=2a,~U(2a)=-\epsilon,$ and a local maximum at
$\phi_{top}=8\lambda a^5/(8\lambda
a^4+3\epsilon),~U(\phi_{top})=128\lambda^3 a^{12} (2\lambda
a^4+\epsilon)/(8\lambda a^4+3\epsilon)^3$. The validity condition
for the thin-wall approximation is $a^4\lambda/\epsilon\gg 1$, see
\cite{Co}. Assuming the thin wall approximation the bounce
solution with a good accuracy may be written as
\begin{equation}\label{bounce}
\phi_b=a\{1-\tanh(\mu[\rho-R])\},\end{equation} where
$\mu=a\sqrt{\lambda}$ and the bubble radius
$R=2\mu^3/\epsilon\lambda$. Following the standard formalism it is
convenient to use the coordinates which respect the symmetry of
the background solution $\phi_b$, i.e. $O(4)$ and $O(3,1)$ in
Euclidean and Minkowskian regions respectively. The coordinate
system in the Euclidean region is given by
$(\rho,\chi,\theta,\varphi)$ where $(\theta,\varphi)$ are usual
angle coordinates on two-dimensional sphere and $(\rho,\chi)$ are
related to $r=|\vec{x}|$ and $\tau$ by
\begin{eqnarray}r=\rho\sin(\chi),~\tau=-\rho\cos(\chi),\nonumber \\0\leq\chi\leq
\pi/2,~0\leq\rho<\infty.\end{eqnarray} The coordinates in the
Minkowskian region are obtained by the replacement $(\rho,\chi)
\rightarrow \\(-i\xi,-i\chi_{M})$, which yields
\begin{eqnarray}r=\xi\sinh(\chi_{M}),~t=\xi\cosh(\chi_{M}),\nonumber
\\0<\chi_{M}<\infty,~0<\xi<\infty.\end{eqnarray} The equation governing the
fluctuation field $\Phi$ reads
\begin{equation}\left[-\partial_{\xi}^2-\frac{3}{\xi}\partial_{\xi}+\frac{1}{\xi^2}\hat{L}^2-U''(\phi_b)\right]\Phi=0,\label{1}\end{equation}
where $\partial_{\xi}$ denotes the partial derivative with respect
to $\xi$ and $\hat{L}$ is the Laplacian operator on
three-dimensional unit hyperboloid. Expanding $\Phi$ in terms of
harmonic functions on the three-dimensional hyperboloid $Y_{plm}$,
$-\hat{L}^2Y_{plm}=(1+p^2)Y_{plm}$, the Eq.(\ref{1}) for the mode
function takes the form
\begin{equation}\label{al}\left[\partial_{\xi}^2+\frac{(p^2+1/4)}{\xi^2}+U''(\phi_{b})\right]f=0.\end{equation}
The analytic continuation of $f$ to the Euclidean region is
performed by setting $\xi\rightarrow i\rho$. Under this
continuation one obtains the equation for the mode function in the
under-barrier region
\begin{equation}\label{fo}\left[\partial_{\rho}^2+\frac{(p^2+1/4)}{\rho^2}-U''(\phi_{b})\right]g=0,\end{equation}
 where $g$ is related to $f$ by the asymptotic
boundary condition $f(\xi)=g(-i\xi)$ at $\xi\rightarrow 0$. At
$\tau\rightarrow -\infty$ the field $\phi$ is in false vacuum
state and correspondingly the fluctuation field satisfies the
vanishing boundary condition when $\rho\rightarrow \infty$. Since
$U''(\phi_b)\rightarrow 4a^2\lambda$ when $\rho\rightarrow\infty$
the asymptotic solution to Eq.(\ref{fo}) satisfying the vanishing
boundary condition is given by
\[g(\rho)=a(p)\sqrt{\rho}K_{ip}(2a\sqrt{\lambda}\rho),\] where
$K_{ip}(x)$ is the modified Bessel function of the second kind. On
the other hand, the asymptotic solution to Eq.(\ref{al}) when
$\xi\rightarrow\infty$ will have the form
\[f(\xi)=c_1(p)e^{-p\pi/2}\sqrt{\xi}\,H^{(1)}_{ip}(2a\sqrt{\lambda}\xi)+
c_2(p)e^{p\pi/2}\sqrt{\xi}\,H^{(2)}_{ip}(2a\sqrt{\lambda}\xi),\]
where $H^{(1)}_{ip}(x)$ and $H^{(2)}_{ip}(x)$ are the Hankel
functions of the first and second kinds, respectively. The
spectrum of created particles $n(p)$ is given by
\begin{equation}\label{sp}n(p)=\frac{1}{||c_1(p)/c_2(p)|^2-1|}.\end{equation}
For detailed description of the general formalism concerning the
problem see \cite{YTS,HSTY}. In what follows, we shall assume that
the particle production does not occur at the moment of bubble
nucleation. To judge the accuracy of this assumption we first
evaluate the spectrum of cretaed particles at the moment of bubble
nucleation. Using the bounce solution (\ref{bounce}) for
$U''(\phi_b)$ one obtains
\[U''(\phi_b)=2\mu^2\{3\tanh^2(\mu[\rho-R])-1\}
+\epsilon\mbox{-term}.\] For the sake of simplicity we approximate
the well of $U''(\phi_b)$ in the region $|\rho-R|<\mu^{-1}$ by the
square well
\begin{equation} U''(\phi_b)=\left\{\begin{array}{ll} 4\mu^2, &\mbox{$|\rho-R|\geq\mu^{-1},$}\\
-d\mu^2,
&\mbox{$|\rho-R|<\mu^{-1},$}\end{array}\right.\end{equation} where
for a reasonable approximation we assume $1.5\leq d<2$. We neglect
the contribution coming from the $\epsilon$-term because at
$\phi=0$ this term equals zero and at $\phi\sim a$ this is of
order $\epsilon\lambda/\mu^2\ll\mu^2$. The solution of
Eq.(\ref{fo}) is given by
\begin{equation} g(\rho)=\left\{\begin{array}{ll} a_1\sqrt{\rho}I_{ip}(2\mu\rho)+a_2\sqrt{\rho}K_{ip}(2\mu\rho), &\mbox{$\rho\leq R_-,$}\\
a_3\sqrt{\rho}H^{(1)}_{ip}(\sqrt{d}\mu\rho)+a_4\sqrt{\rho}\overline{H^{(1)}_{ip}(\sqrt{d}\mu\rho)}, &\mbox{$R_-<\rho<R_+,$}\\
\sqrt{\rho}K_{ip}(2\mu\rho), &\mbox{$\rho\geq
R_+,$}\end{array}\right.\end{equation} where
$R_{\pm}=R\pm\mu^{-1}$, overbar denotes complex conjugation,
$\overline{H^{(1)}_{ip}(x)}=\exp(p\pi)H^{(2)}_{ip}(x)$, and
$I_{ip}(x)$ is the modified Bessel function of the first kind.
Since in the thin-wall approximation $\mu R_{\pm}\gg 1$ for
matching of $g$ at the junction points $R_{\pm}$ one can use the
asymptotic formulas. For $x\gg 1$ and $p^2\ll x$ the asymptotic
expressions are given by \cite{Dun}
\begin{equation}\label{as1} I_{ip}(x)\simeq \frac{1}{\sqrt{2\pi x}}\,e^x,~K_{ip}(x)\simeq
\sqrt{\frac{\pi}{2x}}\,e^{-x},~H^{(1)}_{ip}(x)\simeq\sqrt{\frac{2}{\pi
x}}\,e^ {i(x-ip/2-\pi/4)}.\end{equation} For the coefficients
$a_1,~a_2,~a_3,~a_4$ one obtains
\[a_3\propto (2-i\sqrt{d})\,e^{ -i(\sqrt{d}\mu
R_+-\pi/4)},~a_4\propto -(2+i\sqrt{d})\,e^{i(\sqrt{d}\mu
R_+-\pi/4)},\] where the proportionality coefficients are the same
for $a_3,~a_4$ and
\[a_1\propto \sqrt{\pi}\,(4+d)\sin(2\sqrt{d})e^{-2\mu R_-},~a_2\propto \left\{(4-d)\sin(2\sqrt{d})+
4\sqrt{d}\cos(2\sqrt{d})\right\}e^{2\mu R_-}/\sqrt{\pi},\] the
proportionality coefficients are the same for $a_1,~a_2$. Then for
small values of momentum the number of created particles
$\tilde{n}(p)$ at the moment of bubble nucleation is evaluated as
\begin{equation}
\tilde{n}(p)\simeq\frac{(4+d)^2}{(4\sqrt{d}\cot(2\sqrt{d})+4-d)^2}\exp(-8\mu
R_- ).\end{equation} For large values of momentum $\tilde{n}(p)$
decreases at least as \cite{Ma1}
\begin{equation}\label{nuclspec}\tilde{n}(p)\sim\exp(-2p\pi).\end{equation}

For evaluating the spectrum of created particles during the bubble
expansion one has to solve Eq.(\ref{al}) with the asymptotic
boundary condition $f(\xi)=g(-i\xi)$ at $\xi\rightarrow 0$. The
assumption that the particle production does not occur at the
moment of bubble nucleation implies that the solution to
Eq.(\ref{al}) in the region $\xi<R_-$ is given by \cite{YTS}
\[f(\xi)=c(p)\sqrt{\xi}H^{(1)}_{ip}(2\mu\xi).\] Under this
assumption the solution to Eq.(\ref{al}) reads
\begin{equation} f(\xi)=\left\{\begin{array}{ll} c_1\sqrt{\xi}H^{(1)}_{ip}(2\mu\xi)+c_2\sqrt{\xi}\overline{H^{(1)}_{ip}(2\mu\xi)}, &\mbox{$\xi\geq R_+,$}\\
c_3\sqrt{\xi}L_{ip}(\sqrt{d}\mu\xi)+c_4\sqrt{\xi}K_{ip}(\sqrt{d}\mu\xi), &\mbox{$R_-<\xi<R_+,$}\\
\sqrt{\xi}H^{(1)}_{ip}(2\mu\xi), &\mbox{$\xi\leq
R_-,$}\end{array}\right.\end{equation} where
$L_{ip}(x)=\frac{1}{2}\{I_{ip}(x)+I_{-ip}(x)\}.$ Both $K_{ip}(x)$
and $L_{ip}(x)$ are real valued and even functions of $p$. Any of
$L_{ip}(x),K_{ip}(x)$ and $I_{ip}(x)$ can be constructed from the
remaining two functions by the identity
\[I_{ip}(x)=L_{ip}(x)-i\frac{\sinh(p\,\pi)}{\pi}K_{ip}(x).\]Using
the asymptotic formulas (\ref{as1}) the matching of $f$ at the
junction points $R_{\pm}$ gives \[c_3\propto -\exp(-\sqrt{d}\mu
R_-)(\sqrt{d}+2i)\sqrt{\pi},~c_4\propto \exp(\sqrt{d}\mu
R_-)(2i-\sqrt{d})/\sqrt{\pi},\] where the proportionality
coefficients are the same for $c_3,~c_4$ and \[c_1\propto
i4\sqrt{d}\cosh(2\sqrt{d})+(d-4)\sinh(2\sqrt{d}),~c_2\propto
-(d+4)\sinh(2\sqrt{d}),\] again the proportionality coefficients
are the same for $c_1,~c_2$. Thus the spectrum of created
particles during the bubble expansion for small values of momentum
is approximately given by
\begin{equation}\label{exspec}n(p)\simeq\frac{(d+4)^2}{16d}\sinh^2(2\sqrt{d}).\end{equation}
For large values of $p$ and $x$ one can use the following
asymptotic formulae \cite{Dun}
\[K_{ip}(x)\simeq\frac{\sqrt{2\pi}\,e^{-p\pi/2}}
{(p^2-x^2)^{1/4}}\,\left\{\cos(\alpha(x)-\pi/4)+\frac{1}{12p}\sin(\alpha(x)-\pi/4)\right\},\]\begin{equation}\label{as2}L_{ip}(x)\simeq
-\frac{e^{p\pi/2}}
{\sqrt{2\pi}\,(p^2-x^2)^{1/4}}\left\{\sin(\alpha(x)-\pi/4)-\frac{1}{12p}\cos(\alpha(x)-\pi/4)\right\},\end{equation}
\[H_{ip}^{(1)}(x)\simeq\sqrt{\frac{2}{\pi}}\frac{e^{p\pi/2}}{(p^2+x^2)^{1/4}}\,e^{i(\beta(x)-\pi/4)}\left\{1+\frac{i}{12p}\right\},\]
where $\alpha(x)\equiv
p\ln\{(p+(p^2-x^2)^{1/2})/x\}-(p^2-x^2)^{1/2}$ and $\beta(x)\equiv
p\ln\{x/(p+(p^2+x^2)^{1/2})\}+(p^2+x^2)^{1/2}$. Assume $p\gg\mu R$
then the matching at the junction point $R_-$ gives
\[c_3\propto (2\pi)^{1/2}e^{-p\pi/2}\left\{\frac{p}{\sqrt{d}\mu R_-}[\sin(\gamma)-i\cos(\gamma)]-\frac{\mu
R_-}{2p}\left[\sqrt{d}\sin(\gamma)+\frac{4i}{\sqrt{d}}\cos(\gamma)\right]
\right\},\]\[c_4\propto
-(2\pi)^{-1/2}e^{p\pi/2}\left\{\frac{p}{\sqrt{d}\mu
R_-}[\cos(\gamma)+i\sin(\gamma)]+\frac{\mu
R_-}{2p}\left[\frac{4i}{\sqrt{d}}\sin(\gamma)-\sqrt{d}\cos(\gamma)\right]\right\},\]
where the proportionality coefficients are the same for
$c_3,~c_4$, and $\gamma\equiv\alpha(\sqrt{d}\mu R_-)-\pi/4$. Using
these coefficients and asymptotic formulae (\ref{as2}) from
matching at $R_+$ one obtains that $|c_1/c_2|$ behaves as
$p^2/(\mu R)^2$. Correspondingly, for large values of momentum
\begin{equation}\label{powlow}n(p)\sim \left(\frac{p}{\mu R}\right)^{-4}.\end{equation}
Thus, in contrast to Eq.(\ref{nuclspec}) the particle production
by the expanding thin-walled bubble follows a power law in $p$
when $p\rightarrow\infty$. From Eq.(\ref{powlow}) one concludes
that the contribution to the number of created particles coming
from the ultraviolet region is finite. From Eq.(\ref{exspec}) one
sees that even in the thin-wall case the particle creation is not
really suppressed if the expansion of the bubble is considered.
However, in this case the particle production is not so intensive
as in the thick-wall approximation \cite{Ma2}. This result is not
strictly correct since we neglected the constraint on the particle
production which arises due to fact that the proper fluctuation
field associated with the tunneling is the transverse part of
total fluctuation field with respect to the bounce solution
\cite{Ma3}. Summarizing the above results one concludes that in
the thin-wall approximation particle production is dominated by
the expanding bubble.
\section*{Acknowledgments}
It is our pleasure to acknowledge helpful conversations with
Professors A.\,Khelashvili and I.\,Lomidze.

\end{document}